\documentclass[onecollarge,natbib]{svjour2}
\bibpunct{[}{]}{;}{n}{}{,} 
\smartqed  
\usepackage{graphicx}
\usepackage[utf8]{inputenc}
\usepackage{amsmath}
\usepackage{amssymb}

\newcommand{\einviertel}{ \tfrac{1}{4}}

\journalname{Few-Body Systems}
\begin{document}

\title{Limit Cycles from the Similarity Renormalization Group
}
\subtitle{}


\author{P. Niemann         \and
        H.-W. Hammer 
}


\institute{
Institut für Kernphysik, Technische Universität Darmstadt,
64289 Darmstadt, Germany\\
ExtreMe Matter Institute EMMI,
GSI Helmholtzzentrum für Schwerionenforschung GmbH, 
64291 Darmstadt, Germany\\
}

\date{Received: date / Accepted: date}

\maketitle

\begin{abstract}
We investigate renormalization group limit
cycles within the similarity renormalization group (SRG) and discuss
their signatures in the evolved interaction. 
A quantitative method to detect limit cycles in the interaction
and to extract their period is proposed. Several SRG generators 
are compared regarding their suitability for this purpose. 
As a test case, we consider the limit cycle of the inverse square
potential.
\end{abstract}

\section{Introduction}
Few-body systems with resonant interactions have universal
properties independent of the details of the interaction at
short distances \cite{Braaten:2004rn}.  If the scattering
length $a$ is much larger than the range of the interaction
$r_0$, the Efimov effect can occur \cite{Efimov-70}:
there is a geometric spectrum of three-body bound states
with an accumulation of states at threshold. For identical
bosons in the unitary limit, the Efimov spectrum satisfies the 
geometric scaling relation
\begin{eqnarray}
  \frac{E^{(n)}_3}{E^{(n+1)}_3} = \left(e^{\pi/s_0}\right)^2 \,,
  \label{kappa-star}
\end{eqnarray}
where $s_0=1.00624\ldots$ is a transcendental number.
In general, the value of $s_0$ depends on the symmetries
of the three-body system, the number of interacting pairs,
and the masses of the particles involved, but the form of the
spectrum (\ref{kappa-star}) is universal.
The invariance of the spectrum under discrete scale transformations 
with the preferred scaling factor squared $e^{2\pi/s_0}\approx (22.7)^2$ 
can be understood as the consequence of a three-body 
interaction $H$ governed by a renormalization group 
(RG) limit cycle \cite{Albe-81,Bedaque:1998kg,Mohr:2005pv}. 
The coupling constant
$H(\Lambda)$ is then a periodic function of $\ln(\Lambda)$
with period $\pi/s_0$ where $\Lambda$ is an ultraviolet cutoff
in momentum space used to regulate the theory at short distances. 
More generally, 
the limit cycle will be manifest in observables through their 
log-periodic dependence on the scattering length or other
control parameters of the system \cite{Braaten:2004rn}. 
The period of this log-periodic behavior is again determined 
by the discrete scaling factor which is the key quantity 
determining the properties of the limit cycle.

The universal Efimov spectrum and related higher-body bound states 
have been observed in ultracold atomic
gases in a variety of experiments with different atom
species \cite{Ferlaino:2010, Chin:2010aa, Zenesini:2012}.
In heteronuclear mixtures the scaling factors can be significantly
smaller than for ideal bosons.
For example, in a mixture of $^6$Li and  $^{133}$Cs
atoms, the ratio of subsequent  $^{133}$Cs-$^{133}$Cs-$^6$Li 
bound states is predicted to be $(4.88)^2$. Such heteronuclear
Efimov states have recently been observed in experiment and the 
prediction of the scaling factor was confirmed \cite{Tung:2014,Pires:2014}.
Finally, we note that the Efimov effect also provides a universal 
binding mechanism for S-wave states in weakly-bound nuclei. However, an
observation of the discrete scaling relation between different states
in one nucleus has proven elusive to date since the known S-wave halo nuclei 
have no excited states. (See Ref.~\cite{Hammer:2010kp} for a review.)

Here, we investigate the manifestation of RG limit cycles in the
similarity renormalization group (SRG). The SRG 
was developed independently
by Wegner \cite{Wegner:1994} and by Glazek and Wilson \cite{Glazek:1993rc}.
In recent years, it has become a standard
tool to soften nuclear interactions for improved convergence
in many-body calculations \cite{Bogner:2009bt,Furnstahl:2013oba}. 
The SRG generates a continuous series of unitary transformations 
of the Hamiltonian from the evolution with a flow parameter $s$. 
As the RG flow evolves, the interaction is typically softened at
the expense of introducing induced many-body forces. Ideally, the evolution is 
carried out to a value $s$ such that the interaction is softend enough 
to achieve convergence in many-body calculations without generating 
large contributions from induced many-body forces.
Recent advances in SRG technology have, e.g., 
allowed to extend ab initio calculations
to P-shell nuclei \cite{Jurgenson:2013yya} or perturbative
neutron matter calculations with consistent three-body forces
\cite{Hebeler:2013ri}.
The success of SRG methods in nuclear physics motivates our study of 
the manifestation of RG limit cycles within the SRG.

The emphasis of this work is on formulating criteria for detecting a
limit cycle in the evolved interaction.\footnote{We note that a
limit cycle will also be manifest in physical observables
through universal scaling relations \cite{Braaten:2004rn}.}
In particular, we study the numerical extraction of the discrete scaling 
factor. A long term goal of our work is the derivation of the 
limit cycle in the pionless EFT for the nuclear three-body system.
While it is generally assumed that the pionless EFT
in which the Efimov effect and corresponding limit cycle are manifest
appears as the low-energy limit of a more fundamental chiral EFT with 
explicit pions, no explicit derivation has ever been given.
Our work provides a first step towards this goal.
As a test case, we investigate the attractive inverse square
potential which is known to have an exact limit cycle beyond a critical
coupling strength. For convenience,
we will use natural units with $\hbar=m=1$ in the following.


\section{SRG Basics}
We start with a brief review of the basic properties of the SRG.
A more detailed discussion can be found in 
Refs.~\cite{Furnstahl:2013oba,Kehrein:2006}.

The SRG generates  a continuous series of unitary transformations 
on the Hamiltonian governed by a flow parameter $s$:
\begin{equation}
H(s)=U(s) H(s=0) U(s)^{\dagger}=T +V(s) \; ,
\end{equation}
where $U(s)$ is a unitary operator. Often the kinetic
energy is left unchanged per definition, 
such that the SRG generates an evolution
of the interaction potential $V$. Defining the anti-Hermitian operator
$\eta(s) = (\frac{d}{ds}U(s))U(s)^\dagger = [G(s),H(s)]$, the evolution 
equation for the Hamiltonian can be written as
\begin{equation}
\frac{dH(s)}{ds} = \left[[G(s),H(s)], H(s)\right] \; ,
\label{SRG_1} 
\end{equation}
where the Hermitian operator $G(s)$ is called the generator of the 
SRG transformation. The generator $G$ is often taken independent
of $s$. A frequent
choice for $G$ is the kinetic energy $T$. In this case, the differential 
equation can be written as
\begin{equation}
\frac{dV}{ds}= 2TVT-VTT-TTV+TVV+VVT-2VTV \; .
\end{equation}
If we consider a two-body system with identical particles, 
we can write the flow equation for the two-body interaction $V_2$
in the space of relative momenta as
\begin{equation}
\frac{d}{ds}\langle p|V_2|q \rangle=-(p^2-q^2)^2 \, \langle p| V_2| q \rangle 
+ \int_0^{\infty}\frac{d^3k}{(2\pi)^3}\, (p^2+q^2-2k^2) \langle p|V_2|k \rangle 
\langle k|V_2|q \rangle \;.  
\label{SRG_2} 
\end{equation}
The evolution of partial waves decouple in the two-body system, thus 
Eq.~(\ref{SRG_2}) holds for every partial wave. In this representation 
one can recognize a  major characteristic of the SRG with the $T$
generator. Clearly, the SRG transformation has a fixed point if
$H(s)$ commutes with $T$, i.e. $H(s)$ is diagonal. 

For weak interactions,
the second term in Eq.~(\ref{SRG_2}) can be neglected and the solution
is simply
\begin{equation}
\langle p|V_2(s)|q \rangle =  \langle p| V_2(0)| q \rangle 
\exp(-s(p^2-q^2)^2) \;.
\label{SRG_3}
\end{equation}
All non-diagonal matrix elements approach zero during the evolution and 
thus a decoupling between low-energy and high-energy matrix elements is 
achieved.  Although the second term in Eq. (\ref{SRG_2}) typically can 
not be neglected, the suppression of non-diagonal matrix elements 
is a dominant part of the full SRG evolution using the $T$ generator. 
The approximate expression (\ref{SRG_3}) also suggests to
interpret  
\begin{equation}
\label{eq:cut}
\lambda \equiv s^{-1/4}
\end{equation}
as an effective momentum cutoff.
Thus the evolution starts at $s=0$ corresponding to $\lambda =\infty$
and as $s$ increases, the effective momentum cutoff $\lambda$
is lowered.

In this work, we also use two alternative generators, 
which are functions of the kinetic energy. 
They were introduced by Li, Anderson, and Furnstahl
with the aim to obtain a more efficient evolution which is crucial 
for identifying limit cycles  \cite{Li:2011sr}. 
We refer to them as  exponential generator, $G_e$,
and inverse  generator, $G_i$:
\begin{align}
\label{eq:agen}
G_e &= -\sigma^2 \exp(-T/\sigma^2)\,, \notag \\
G_i &=\frac{-\sigma^2}{1+T/ \sigma^2}\,,
\end{align}
where $\sigma$ is an arbitrary parameter with dimensions of momentum.
Both generators have a power series expansion in $T$.
For small momenta $q\ll \sigma$, they approach $T$ up to a constant 
which cancels out in $\eta$. So depending on the 
parameter $\sigma$, there is a separation into a low-energy region, 
where the two generators behave like the $T$ generator and a high-energy 
region, where the SRG evolution is suppressed. 
A detailed discussion of these generators can be found in Ref.~\cite{Li:2011sr}.

Since $\sigma$ has dimensions of momentum, the translation of
$s$ to an effective momentum cutoff is more subtle than for the 
$T$ generator. The solutions of  Eq.~(\ref{SRG_2})  in the weak interaction 
limit are
\begin{equation}
\langle p|V_2(s)|q \rangle =  \langle p| V_2(0)| q \rangle 
\exp[-s\sigma^2 (q^2-p^2)(e^{-p^2/\sigma^2}-e^{-q^2/\sigma^2})]\,, 
\quad \mbox{ for } G_e\,,
\end{equation}
and
\begin{equation}
\langle p|V_2(s)|q \rangle =  \langle p| V_2(0)| q \rangle 
\exp\left[-s\sigma^2 (q^2-p^2)\left(\frac{1}{1+p^2/\sigma^2}-
\frac{1}{1+q^2/\sigma^2}\right)\right]
\quad \mbox{ for } G_i\,.
\end{equation}
Thus the effective momentum cutoff is $\lambda \sim \sigma^{-1}s^{-1/2}$.
The constant $\sigma$ is irrelevant for the 
parametric dependence on $s$. Setting it to one, we define the 
effective momentum cutoff
\begin{equation}
\lambda_a \equiv s^{-1/2}
\end{equation}
for the exponential and inverse generators from Eq.~(\ref{eq:agen}).
This will have important consequences when extracting 
limit cycle periods from the evolved interaction below.

\section{Renormalization of the $1/R^2$-Potential}
In the following, we discuss the quantum mechanical $1/R^2$ potential
as a test case.
This is a singular potential which displays an exact limit cycle. 
We start by reviewing the 
renormalization of the  $1/R^2$ potential in an effective 
field theory framework. Here, the limit cycle becomes manifest in the 
behavior of a counter term. We follow the discussion in 
Ref.~\cite{Hammer:2005sa} where further details can found.
In the next section, we investigate the 
$1/R^2$ potential in the SRG framework and provide general criteria 
for isolating limit cycle behavior in the interaction.

The  $1/R^2$ potential can be written as
\begin{equation}
V(R)=\frac{c}{R^2}
\end{equation}
with $R:=|\mathbf{R}|$ and $c$ a coupling constant. 
For subcritical couplings $c>-\einviertel$, the 
potential is well behaved and leads to a unique solution
of the Schr\"odinger equation. For critical and supercritical values
$c\leq-\einviertel$, however, the potential is singular and displays a limit 
cycle. In this case, it is useful to define a parameter $\nu$
that characterizes the period of the limit cycle via
$\nu:=\sqrt{-c-\einviertel}$,

The momentum-space representation of the potential can be defined 
via a Fourier transform in $D$ dimensions,
\begin{equation}
V(Q)= \lim_{D \rightarrow 3} \int d^D \mathbf{R}\, e^{i \mathbf{Q \cdot R}} 
\,V(R) 
=\frac{2 \pi^2c}{Q} \; ,
\end{equation}
where $Q$ is the momentum transfer~\cite{Hammer:2005sa}.

In the following, we consider only S-waves.
For the momentum space matrix elements of the 
S-wave projected potential, we get
\begin{equation}
V(p,q)=2\pi^2 c \left( \frac{\theta(p-q)}{p} 
+\frac{\theta(q-p)}{q} \right) \; ,
\end{equation}
where $q$ ($p$) are the incoming (outgoing) momenta.
The physical observables can be obtained from the Lippmann-Schwinger 
equation
\begin{equation}
t_E(p,k)=V(p,k) + \frac{1}{2 \pi^2} \int_0^{\Lambda} 
\frac{dq \, q^2}{E-q^2+i \epsilon} V(p,q) t_E(q,k) \; ,
\label{eq:ls-1oRq}
\end{equation}
where $E=k^2$ is the total energy and the scattering phase shifts are
given by
\begin{equation}
k\cot\delta = ik-\frac{4\pi}{\left.t_E (k,k)\right|_{E=k^2}}\,.
\end{equation}
The bound states are given by the solutions of the corresponding
homogeneous equation.
As discussed in \cite{Hammer:2005sa}, Eq.~(\ref{eq:ls-1oRq}) 
has no unique solution 
for $\Lambda \to \infty$ if $c < -\einviertel$
and requires renormalization. We
regulate Eq.~(\ref{eq:ls-1oRq}) with a sharp momentum cutoff $\Lambda$
and absorb the cutoff dependence by introducing a momentum independent 
counterterm $\delta V(\Lambda)$.
\begin{equation}
V(p,q) \Rightarrow V(p,q)+\delta V(\Lambda) 
=2\pi^2 c \left( \frac{\theta(p-q)}{p} +\frac{\theta(q-p)}{q}  + \frac{H(\Lambda)}{\Lambda} \right)\,.
\label{eq:VH}
\end{equation}
Demanding invariance of the zero-energy solution under changes 
of $\Lambda$, one finds
\begin{equation}
H(\Lambda)=\frac{1-2 \nu \tan(\nu \ln(\Lambda / \Lambda_*))}
{1+2 \nu \tan(\nu \ln(\Lambda / \Lambda_*))} 
=1-4 \nu^2 \ln(\Lambda / \Lambda_*) + \ldots\; ,
\label{eq:ctexplicit}
\end{equation}
where $\Lambda_*$ is a low-energy constant. Including this counterterm in
Eq.~(\ref{eq:ls-1oRq}) keeps all low-energy observables fixed
when $\Lambda$ is varied.
One can immediately see that the counterterm $H(\Lambda)$ displays a limit 
cycle with a preferred scaling factor $\exp(\pi / \nu)$ since
$\tan$ is a periodic function with period $\pi$.
If the cutoff $\Lambda$ is changed by multiples of  $\exp(\pi / \nu)$,
the counterterm returns to the same value.

The bound state spectrum satisfies a geometrical scaling relation
analog to the Efimov case,
\begin{equation}
\frac{E^{(n)}}{E^{(n+1)}} = e^{2\pi/\nu}\,,
\end{equation}
and presents an ideal test case for the application of SRG methods to 
limit cycles.
In the following section, we will investigate the limit cycle in the inverse
square potential using the SRG framework.
\section{$1/R^2$ Potential and SRG}
\label{1/R2SRG}
In this section, we consider the $1/R^2$ potential in the SRG framework. 
Since the SRG is a unitary transformation, all observables stay constant 
during the evolution  by definition. So in contrast to the explicit 
construction of the counterterm in the effective field theory
treatment of the previous subsection, we need to extract a signal 
for the limit cycle from the evolved interaction.
In order to define such a signal, we investigate the SRG evolution
of the $1/R^2$ potential for critical and subcritical couplings 
and different generators. 

Before we proceed, we specify our units. One free length scale $l_0$ 
is present in our framework. 
Therefore, all dimensionful quantities are given in units of $l_0$. 

\subsection{Qualitative Features}
\begin{figure}
\begin{minipage}[hbt]{8cm}
	\centering
	\includegraphics[width=8cm]{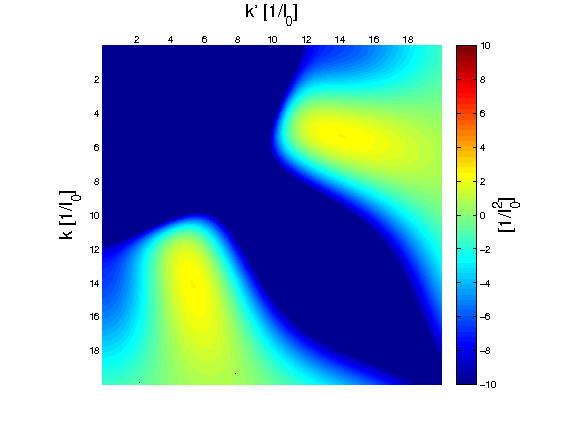}
\end{minipage}
\hfill
\begin{minipage}[hbt]{8cm}
	\centering
	\includegraphics[width=8cm]{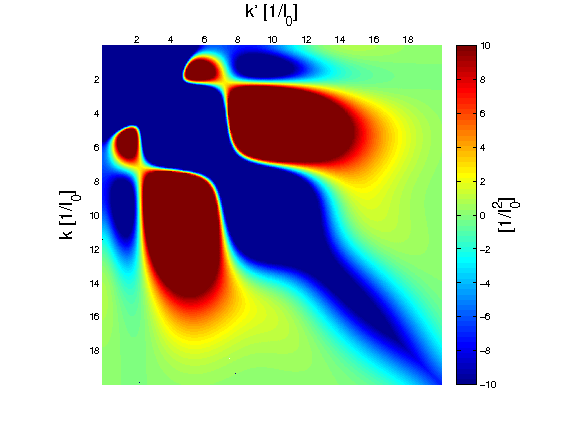}
\end{minipage}\\
\begin{minipage}[hbt]{8cm}
	\centering
	\includegraphics[width=8cm]{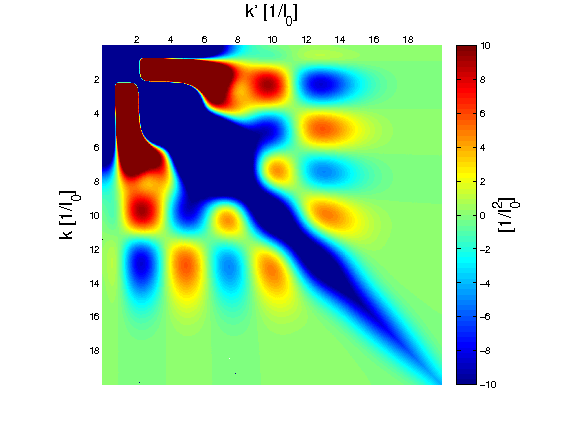}
\end{minipage}
\hfill
\begin{minipage}[hbt]{8cm}
	\centering
	\includegraphics[width=8cm]{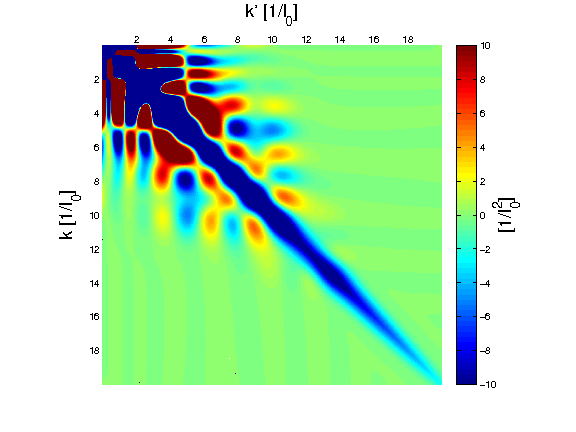}
	\end{minipage} \\
	\begin{minipage}[hbt]{8cm}
	\centering
	\includegraphics[width=8cm]{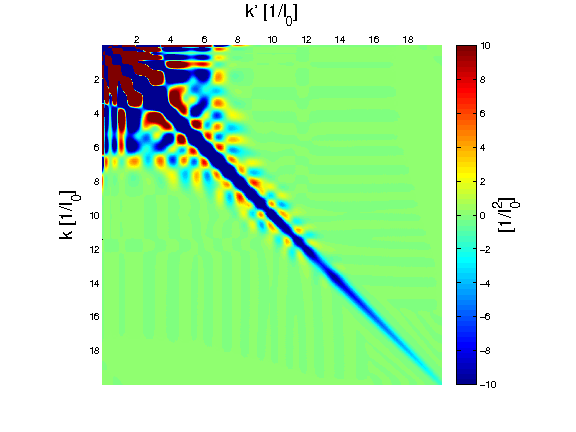}
\end{minipage}
\hfill
\begin{minipage}[hbt]{8cm}
	\centering
	\includegraphics[width=8cm]{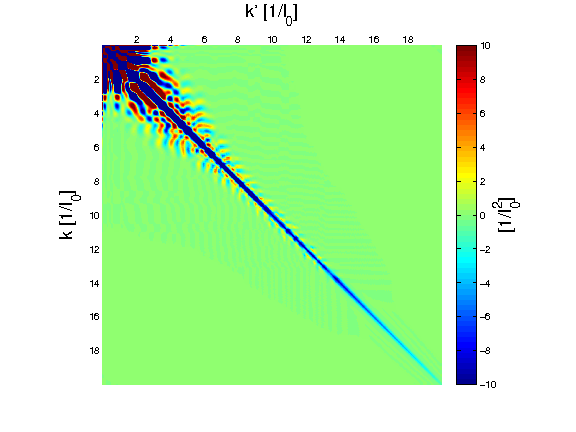}
	\end{minipage} 
	\caption{Evolution of the S-wave $1/R^2$ potential for the 
$T$ generator with $\nu=9$ and initial cutoff $\Lambda=20 \;l_0^{-1}$. 
$V(k,k',\lambda)$ is shown 
for $\lambda \approx$  21.09, 13.67, 8.86, 5.74, 3.72 and  2.41 
in units of $[1/l_0]$ from top left to bottom right.}
\label{fig:T_pictures}
\end{figure}
First, we consider the standard $T$ generator for the SRG 
transformation.  In Fig. \ref{fig:T_pictures},
we show the evolution of the potential for $\nu=9$ and 
an initial cutoff in Eq.~(\ref{eq:ls-1oRq}) of  $\Lambda=20 \;l_0^{-1}$.
Introducing this regulator is required in order to insure
that  Eq.~(\ref{eq:ls-1oRq}) has a unique solution. 

Changing the 
value of $\Lambda$ corresponds to changing the short-distance
behavior of the starting interaction.
We leave $\Lambda$ fixed and investigate the dependence of the 
interaction on the SRG cutoff $\lambda$ that is directly related to the
flow parameter $s$. 
Striking is the appearance of separated regions in the potential 
with positive and negative sign. To make them clearly visible, we choose 
a rather small  maximum value for the coloring of the potential. As
the evolution progresses, these 
regions are constantly emerging and vanishing while the total number 
of regions increases. At the beginning of the evolution, two positive valued 
regions appear.  During the further progress more 
and more positive and negative 
regions emerge. The size of these structures also decreases, which 
is related to the general suppression of off-diagonal matrix elements in the 
SRG evolution for the $T$ generator. 
In the last picture, one can clearly see the large  
number of small regions. We also note that this behavior occurs on a 
logarithmic scale of the flow parameter $s$.

To confirm that the appearance of the oscillatory behavior is indeed
related to the limit cycle, we have evolved $1/R^2$ potentials 
with subcritical coupling $c > -\einviertel$ where no limit cycle 
occurs and critical couplings $c \leq -\einviertel$ where it is 
present. 
\begin{figure}
\begin{minipage}[hbt]{8cm}
	\centering
	\includegraphics[width=8cm]{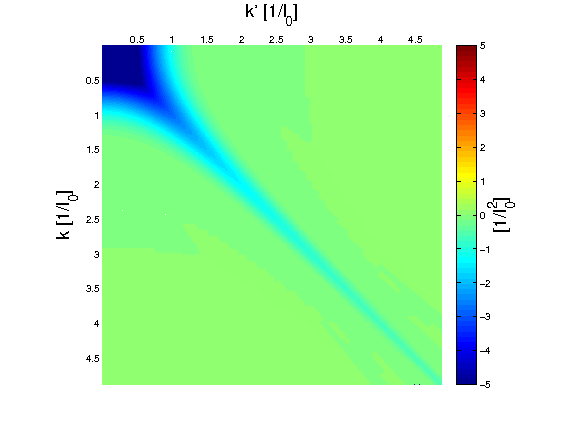}
\end{minipage}
\hfill
\begin{minipage}[hbt]{8cm}
	\centering
	\includegraphics[width=8cm]{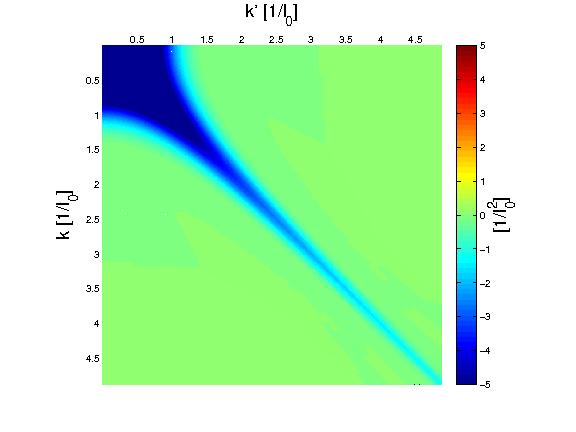}
\end{minipage}\\
\begin{minipage}[hbt]{8cm}
	\centering
	\includegraphics[width=8cm]{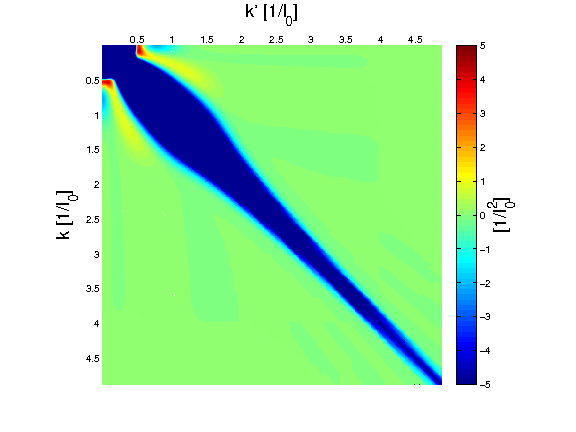}
\end{minipage}
\hfill
\begin{minipage}[hbt]{8cm}
	\centering
	\includegraphics[width=8cm]{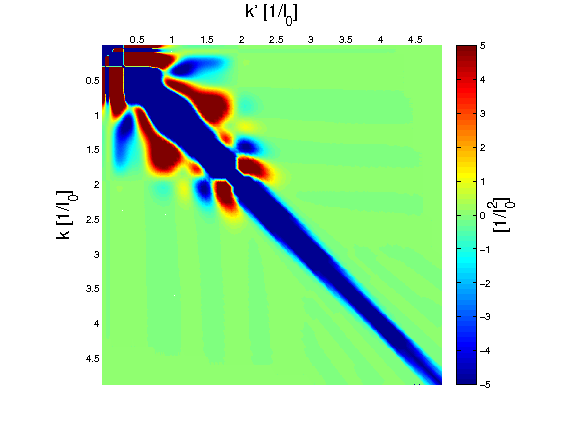}
	\end{minipage} \\
	\caption{S-wave $1/R^2$ potential evolved with the 
$T$ generator to $\lambda=1 \;l_0^{-1}$ with initial $\Lambda=20 \;l_0^{-1}$. 
The potential 
strengths are $c=-0.125$, $c=-0.25$, $c=-0.25 -1^2$ and $c=-0.25 -2^2 $ 
in the order from top left to bottom right.}
\label{fig:T_nuverg}
\end{figure}
In Fig. \ref{fig:T_nuverg}, four potentials are depicted, which were 
all evolved to $\lambda=1 \;l_0^{-1}$. All parameters of the potentials 
except for the coupling constant $c$ are kept constant.
For $c<-\einviertel$, the scaling factor is given by $\exp(\pi / \nu)$
with $\nu =\sqrt{-c-\einviertel}$. 
Thus, if $c$ approaches the critical value $-\einviertel$ 
the scaling factor diverges.
In Fig. \ref{fig:T_nuverg}, the evolved potentials beneath and at the 
critical value do not exhibit the oscillatory behavior. Only the effective
diagonalization of $V$ from the SRG transformation is clearly visible. 
For $\nu=1$, two separated  regions with opposite signs
are observable up to this point in the 
evolution and for $\nu=2$ several structures are already visible. This
observation clearly supports our conjecture that the appearance of the
oscillatory features is related to the limit cycle.

\begin{figure}
\begin{minipage}[hbt]{8cm}
	\centering
	\includegraphics[width=8cm]{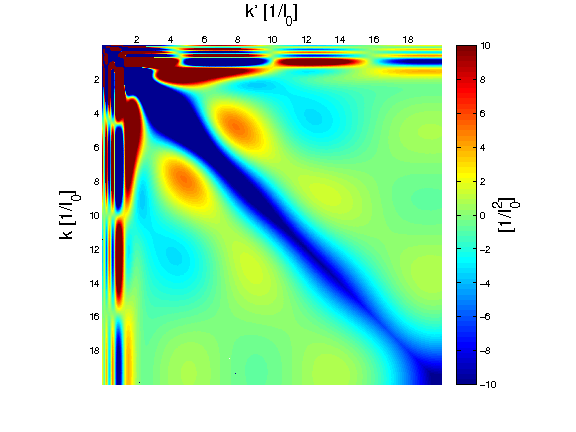}
\end{minipage}
\begin{minipage}[hbt]{8cm}
	\centering
	\includegraphics[clip,width=8cm,angle=0]{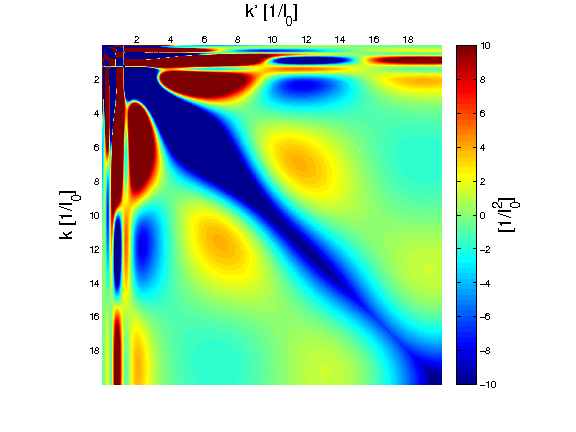}
\end{minipage}\\
\caption{Evolved $1/R^2$ potential with  parameters  $\nu=9$, 
$\Lambda=20 \;l_0^{-1}$ 
and $\sigma=2 \; l_0^{-1}$ for $G_e$ and $G_i$. Left panel: 
Evolution with the exponential 
generator $G_e$ to $\lambda_a \approx 8.35 \;l_0^{-1} $. Right panel: 
Evolution with the 
inverse generator $G_i$ to $\lambda_a \approx 10.75 \;l_0^{-1}$.}

\label{fig:rquad_expT_invT}
\end{figure}

Next, we consider the inverse and exponential generators. 
We expect a similar qualitative signature of the limit cycle.
However, the alternative 
generators contain a free dimensionful parameter $\sigma$ 
which divides the potential into  two different regions. For small momenta
compared to $\sigma$, the exponential and 
inverse generators reduce to the $T$ generator. For large momenta, 
the generators approach zero and
the evolution is suppressed. As an example,
we have plotted the evolved potential for $\nu=9$, $\Lambda=20 \;l_0^{-1}$,
$\sigma=2 \; l_0^{-1}$ in Fig. \ref{fig:rquad_expT_invT}
for both the inverse and the exponential generator.

Notable is the fact that the oscillatory features become 
compressed in a rather small area in the $k-k'$ plane, whose size depends 
on $\sigma$. 
We remark that the qualitative behavior of the exponential and inverse 
generators are very similar. Thus, we will not distinguish their traits 
here.  As for the standard $T$ generator, the oscillatory structures 
only appear if the coupling constant is supercritical.

\subsection{Discrete Scaling Factor}
Next we focus on the question of how to extract the discrete scaling factor
as the main characteristic of the limit cycle from the evolved potential.
To observe a log-periodic signal, we investigated several different
strategies, which we will discuss in the following.

First, we have investigated the possibility to isolate the 
oscillatory feature by projecting on the momentum independent
part of the evolved potential. Assuming that the short-range
part of the evolved potential can be expanded as
\begin{equation}
V(k,k',\lambda)=C_0(\lambda) + \frac{C_2(\lambda)}{2} (k^2+k'^2) + \ldots\,,
\end{equation}
we have investigated the possibility to extract the preferred 
scaling factor from the $\lambda$ dependence of $C_0$.
In particular, we considered the quantities
 \begin{equation}
I_1(\lambda) \equiv \int \int d^3k\, d^3k'\, V(k,k',\lambda) \qquad 
\mbox{ and } \qquad I_2(\lambda) \equiv  \int d^3k\, V(k,k,\lambda) \; .
 \end{equation}
$I_1(\lambda)$ is the projection of the evolved potential 
whereas $I_2(\lambda)$ is the projection of the diagonal part
of the evolved potential. In both cases, we were not able to 
detect any clear signals of the limit cycle.

Second, we examined the diagonal elements of the evolved potential 
$V(p,p,s)$ in dependence of the flow parameter. This is motivated by 
the diagonalizing factor $\exp(-s(p^2-q^2)^2)$ from Eq.~(\ref{SRG_3}). On the 
diagonal of the potential matrix incoming and outgoing momenta are identical, 
so that the exponential function is one. Hence, the diagonal elements are 
the only ones, which do not approach zero during the evolution. So, we 
expect a log-periodic signal to be most prominent on the diagonal.
A similar strategy was followed by Glazek \cite{Glazek:2006mg} in the 
analysis of a discrete model displaying a limit cycle.

\begin{figure}
\begin{minipage}[hbt]{7.5cm}
	\centering
	\includegraphics[width=7.5cm]{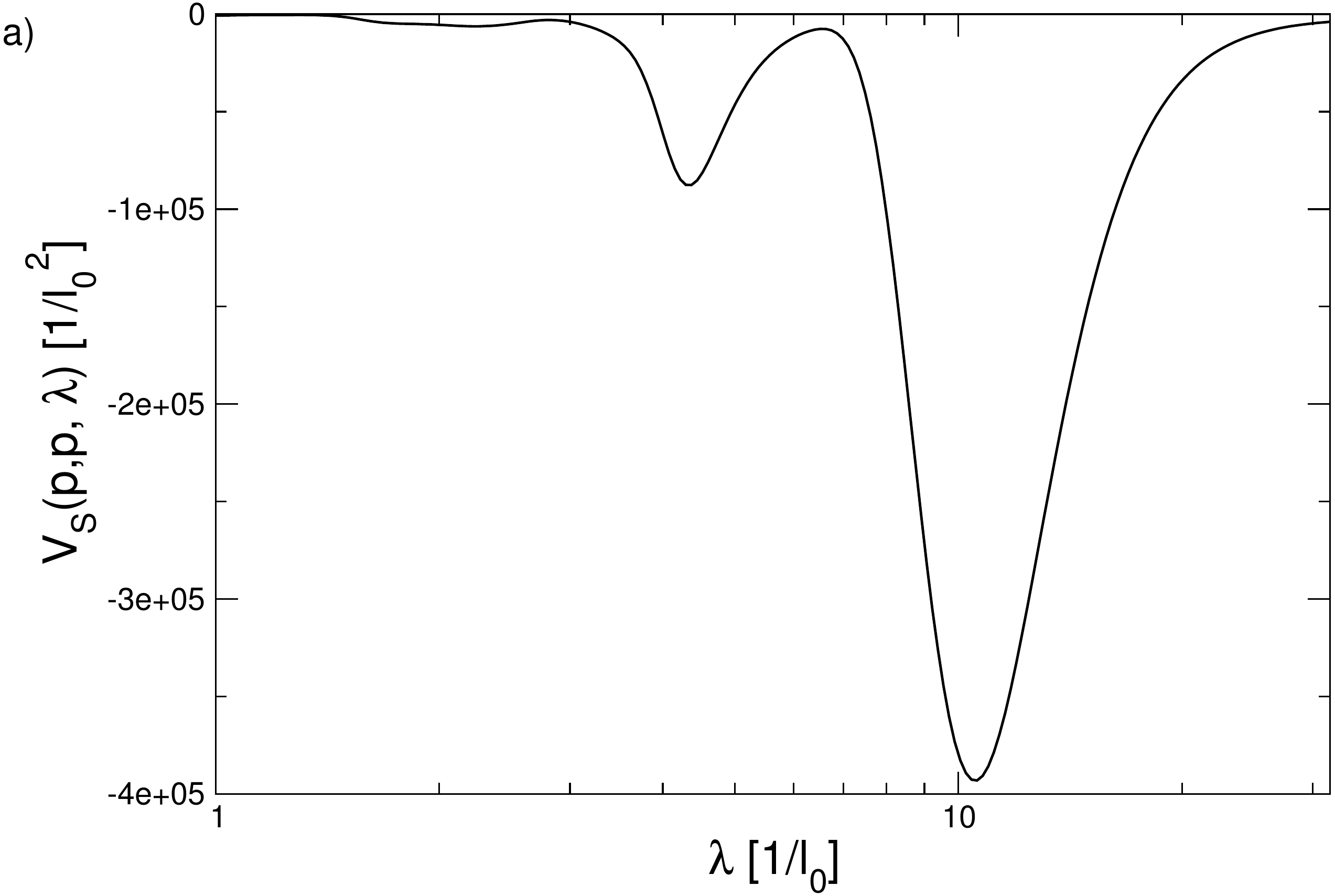}
\end{minipage}
\hfill
\begin{minipage}[hbt]{7.5cm}
	\centering
	\includegraphics[clip,width=7.5cm,angle=0]{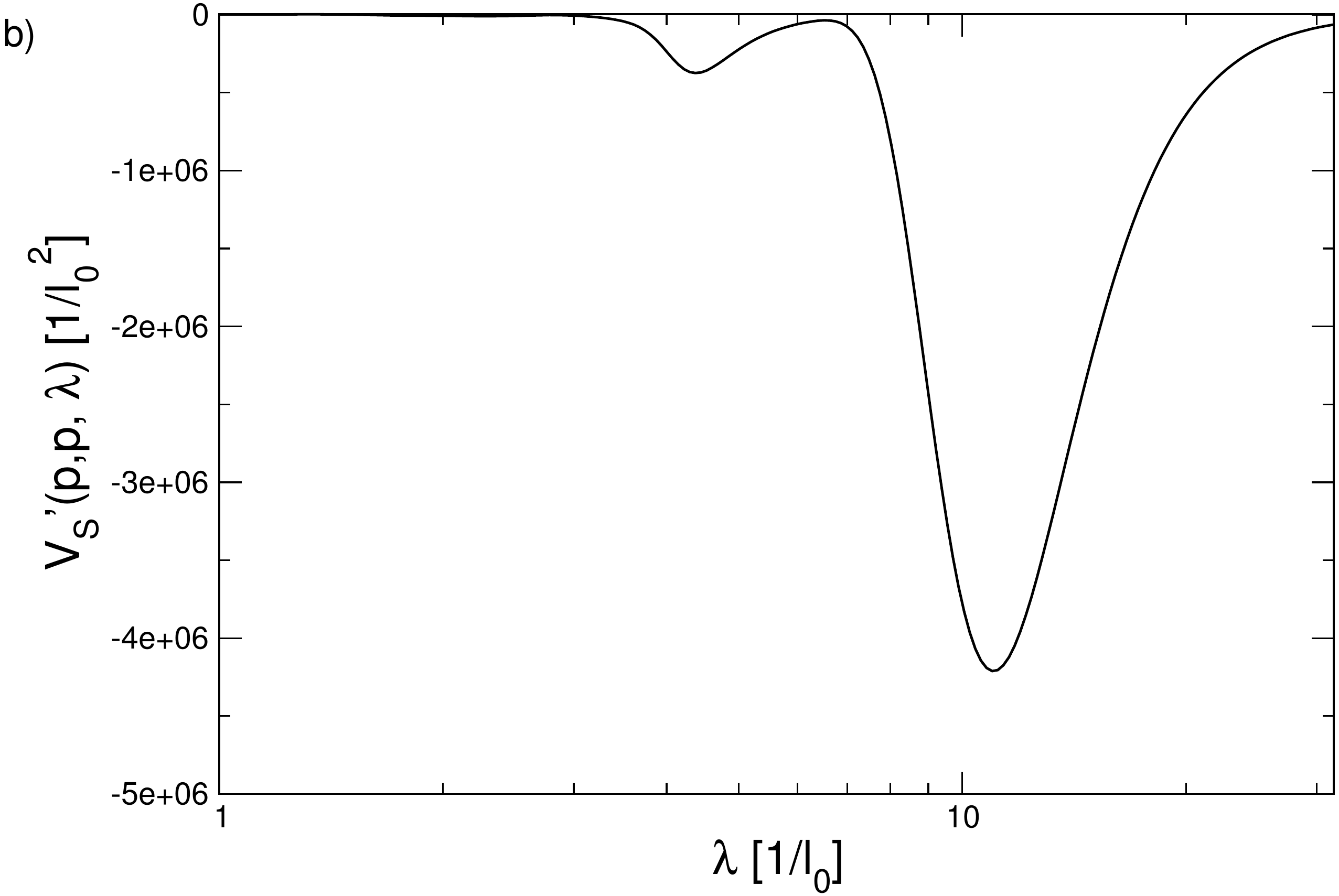}
\end{minipage}\\
\caption{(a) Diagonal element $V(p,p,\lambda)$ in dependence of 
$\lambda$ for $p \approx 0.84 \;l_0^{-1}$. 
(b) $V'(p,p,\lambda)$ from Eq.~(\ref{eq:Vp}) 
in dependence of $\lambda$ for $p \approx 0.84 \;l_0^{-1}$. 
The parameters of the potential are  $\nu=9$ and  $\Lambda=20 \;l_0^{-1}$. 
The evolution was carried out with the standard $T$ generator.}
\label{fig:T_diag}
\end{figure}

{\it $T$ generator.}\quad
We start with the standard $T$ generator.
In Fig. \ref{fig:T_diag}(a) a typical diagonal element is depicted 
in dependence  of the flow parameter. 
The diagonal elements show some irregular oscillations but a clear 
signature of the limit cycle period can not be extracted. This is also
the case if we subtract the initial potential from the evolved potential
and multiply by $\lambda$ 
in order to isolate the SRG analog of the counterterm $H$ 
from Eq.~(\ref{eq:VH}) above:
\begin{equation}
V'(p,p,\lambda)=(V(p,p,\lambda)-V(p,p,\lambda= \infty) ) \cdot 
\lambda \, , \quad\mbox{ where }\quad  \lambda = s^{-1/4}\,.
\label{eq:Vp}
\end{equation}
In Fig.~\ref{fig:T_diag}(b), we show 
$V'(p,p,\lambda)$ from Eq.~(\ref{eq:Vp}) 
for the same diagonal matrix element as in 
Fig. \ref{fig:T_diag}(a).  Again a clear signature of the limit
cycle period could not be extracted. This was also the case if 
different powers of $\lambda$ were used in Eq.~(\ref{eq:Vp}). 

\begin{figure}
\begin{minipage}[hbt]{7.5cm}
	\centering
	\includegraphics[width=7.5cm]{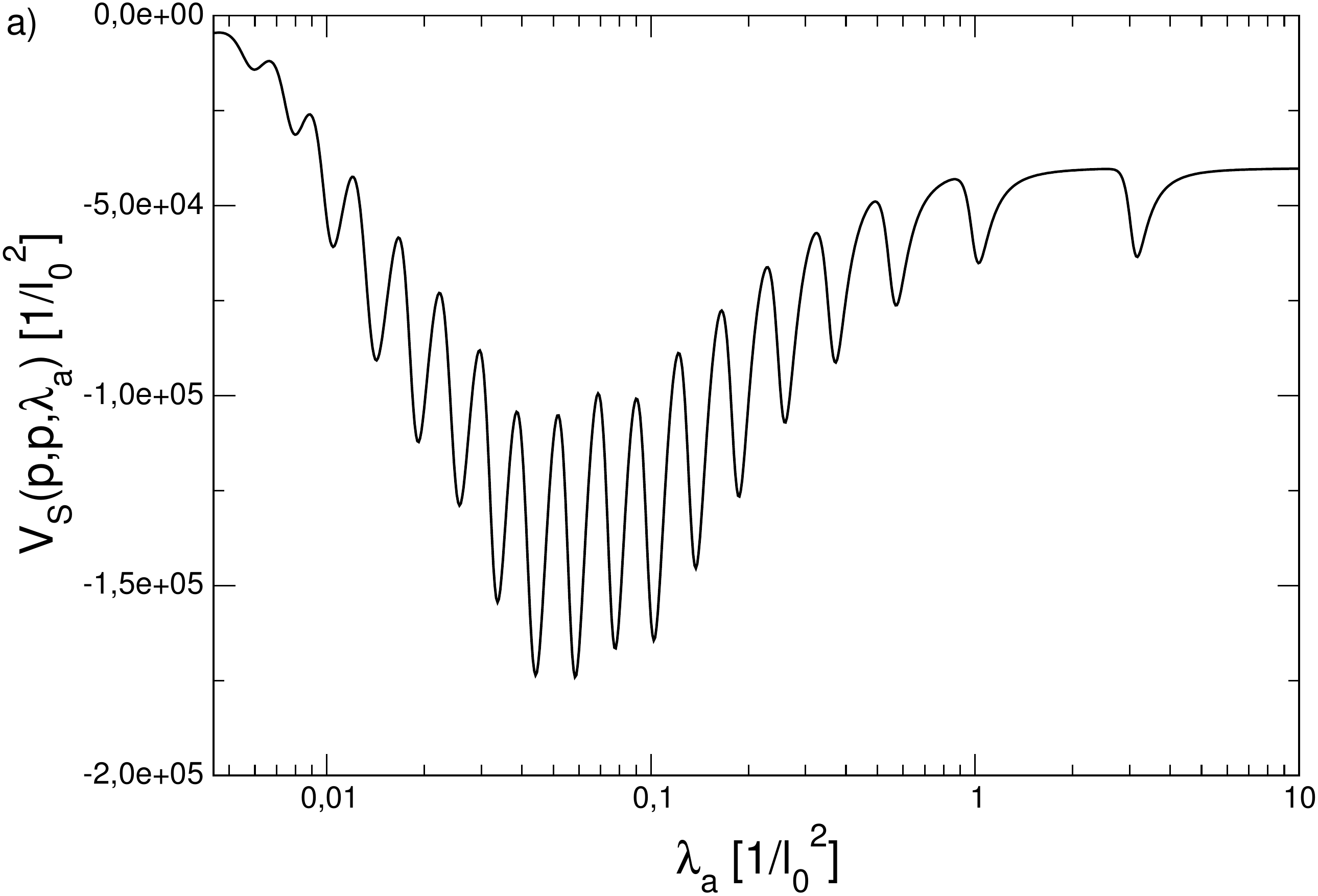}
\end{minipage}
\hfill
\begin{minipage}[hbt]{7.5cm}
	\centering
	\includegraphics[clip,width=7.5cm,angle=0]{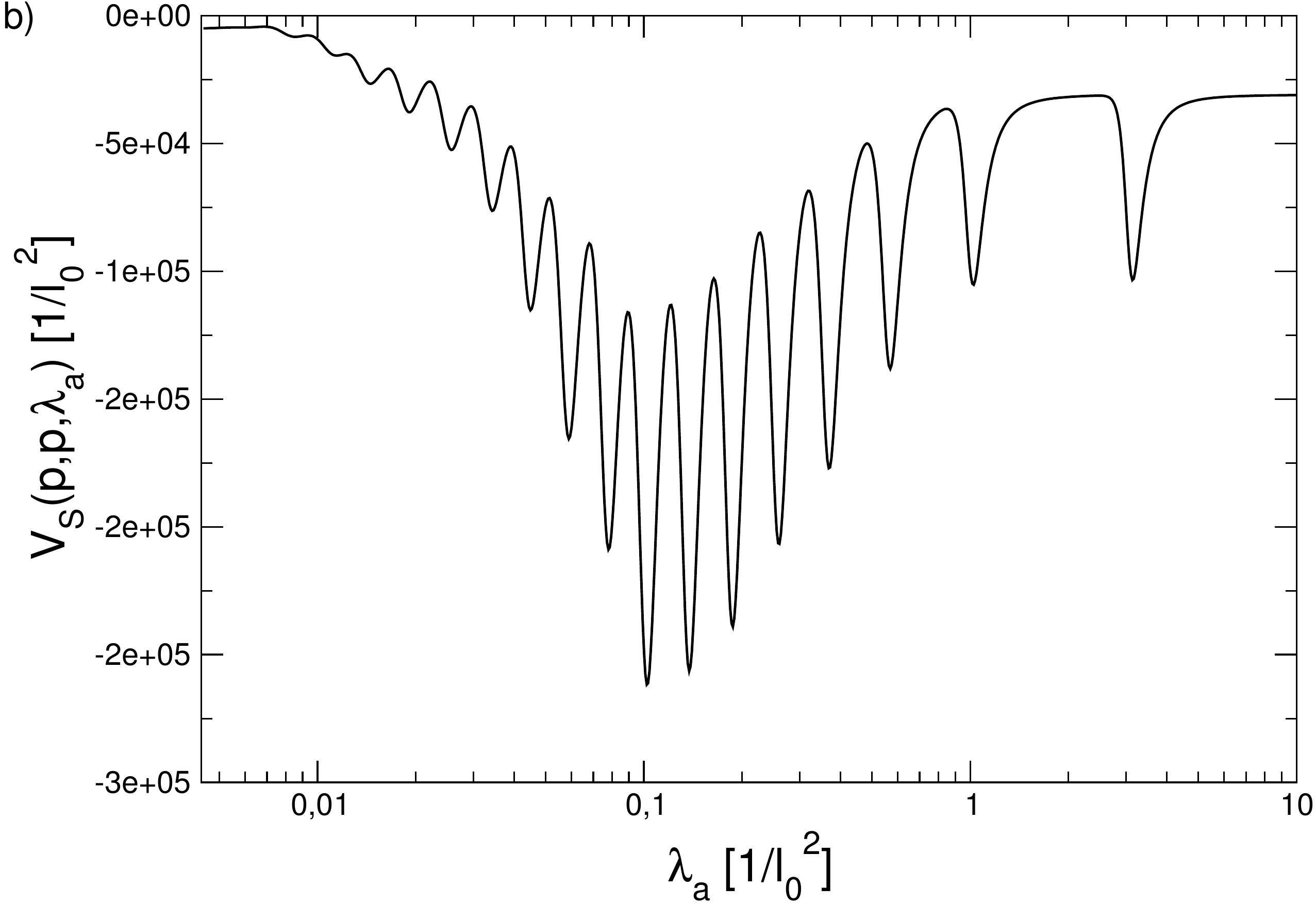}
\end{minipage}\\
\caption{Evolution of the diagonal elements $V(p,p,\lambda_a)$ with $\nu=11$, $\Lambda=30 \;l_0^{-1}$  
and  $\sigma=0.05 \; l_0^{-1}$. (a) Applied exponential generator  with $p \approx 1.19 \sigma$  (b)  
Applied inverse generator evolved with  $p \approx 1.55 \sigma$.}
\label{fig:Diagonale}
\end{figure}
{\it Alternative generators.}\quad
Using the standard $T$ generator only very few irregular
oscillations can be observed. 
Therefore, we try the same strategy with the exponential and 
inverse generators that allow for a further evolution in $s$.
Here, a completely different behavior is found if momenta of
order $\sigma$ are considered. In  Fig.~\ref{fig:Diagonale}, 
we plot a diagonal element with  momentum $p$ close to the parameter 
$\sigma$ in dependence of the flow parameter for both generators. One can 
now clearly see regular oscillations for both generators. The graphs look like 
a log-periodic function multiplieded with another 
slowly-varying function. 

An example with explicit values is given in Table~\ref{tab1}. The extracted 
distances between the maxima and minima are constant to about 5\%, except 
for the first few oscillations at large values of $\lambda_a$ which
is probably caused by finite cutoff effects. 
The period depends on the strength of the 
initial potential $\nu$. Larger values of $\nu$ result in smaller periods. 
We find that the extracted periods are in good agreement
with the exact formula $\exp(\pi/\nu)$. The agreement is better
for larger values of $\nu$, where more oscillations can be seen and
the period can be determined more accurately. 
\begin{table}
\centering
\begin{tabular}{|l|l|l|l|l|}
\hline
oscillation & \multicolumn{2}{|c|}{$\nu=11$} & \multicolumn{2}{|c|}{ $\nu=5$} \\
& maxima & minima  & maxima & minima\\
\hline
 1   & 2.94  &  3.08	&	3.26 &	3.36 \\
 2   & 1.76  &  1.80	&	2.11 &	2.11 \\
 3   & 1.52  &  1.54	&	1.91 &	1.91 \\
 4   & 1.42  &  1.43	&	1.87 &	1.85 \\
 5   & 1.38  &  1.39	&	1.83 &	1.81 \\
 6   & 1.36  &  1.36	&	1.85 &	1.80 \\
 7   & 1.35  &  1.34	&	1.84 &	1.79 \\
 8   & 1.31  &  1.32	& 	     &  1.78 \\
 9   & 1.33  &  1.33	&            &       \\
 10  & 1.34  &  1.33	&            &         \\
 11  & 1.30  &  1.31	&            &		\\
 12  & 1.33  &  1.31	& 	     &		\\
 13  & 1.34  &  1.34	& 	     &		\\
 14  & 1.38  &  1.35	& 	     &		\\
 15  & 1.36  &  1.36	& 	     &		\\
 16  & 1.33  &  1.31	& 	     &    	\\
 17  &       &  1.34	& 	     &    	\\
 \hline
\end{tabular}
\caption{Examples of extracted ratios $\lambda^{(i)}_a / \lambda_a^{(i+1)}$ for the exponential 
generator, where the $\lambda^{(i)}_a$ are the flow parameter values of the 
maxima (minima) of 
$V(p,p,\lambda_a)$ with $p \approx 1.19 \sigma$ for $\nu=11$ 
and $\nu=5$. The parameter values are $\Lambda=30 \;l_0^{-1}$ and $\sigma=0.05 \; l_0^{-1}$.
The exact scaling factors are $\exp(\pi /11) \approx 1.33$ and 
$\exp(\pi / 5) \approx 1.87$.}
\label{tab1}
\end{table}

\begin{figure}
\begin{minipage}[hbt]{7.5cm}
	\centering
	\includegraphics[width=7.5cm]{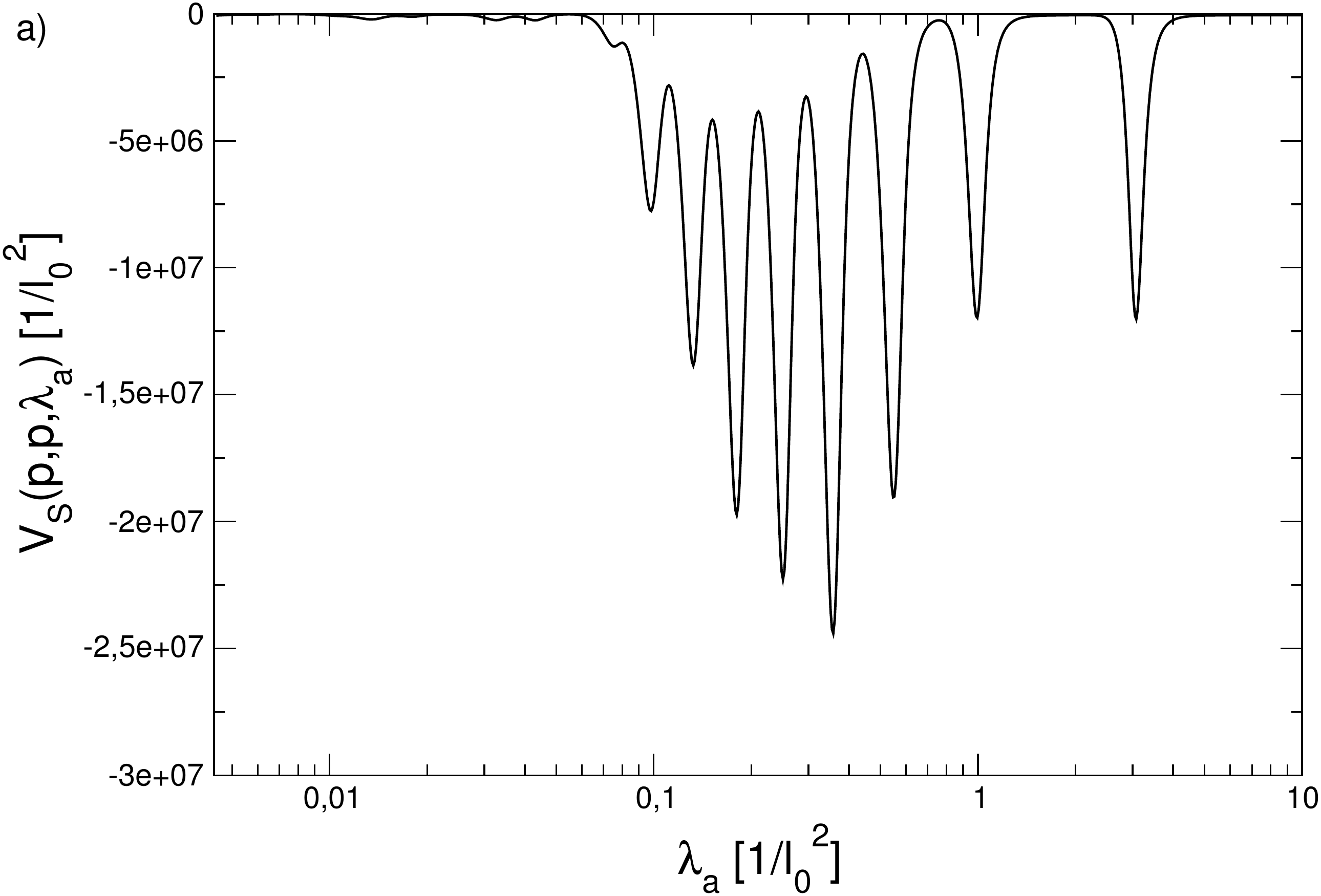}
\end{minipage}
\hfill
\begin{minipage}[hbt]{7.5cm}
	\centering
	\includegraphics[width=7.5cm]{invT_11_b2.pdf}
\end{minipage}\\
\begin{minipage}[hbt]{7.5cm}
	\centering
	\includegraphics[width=7.5cm]{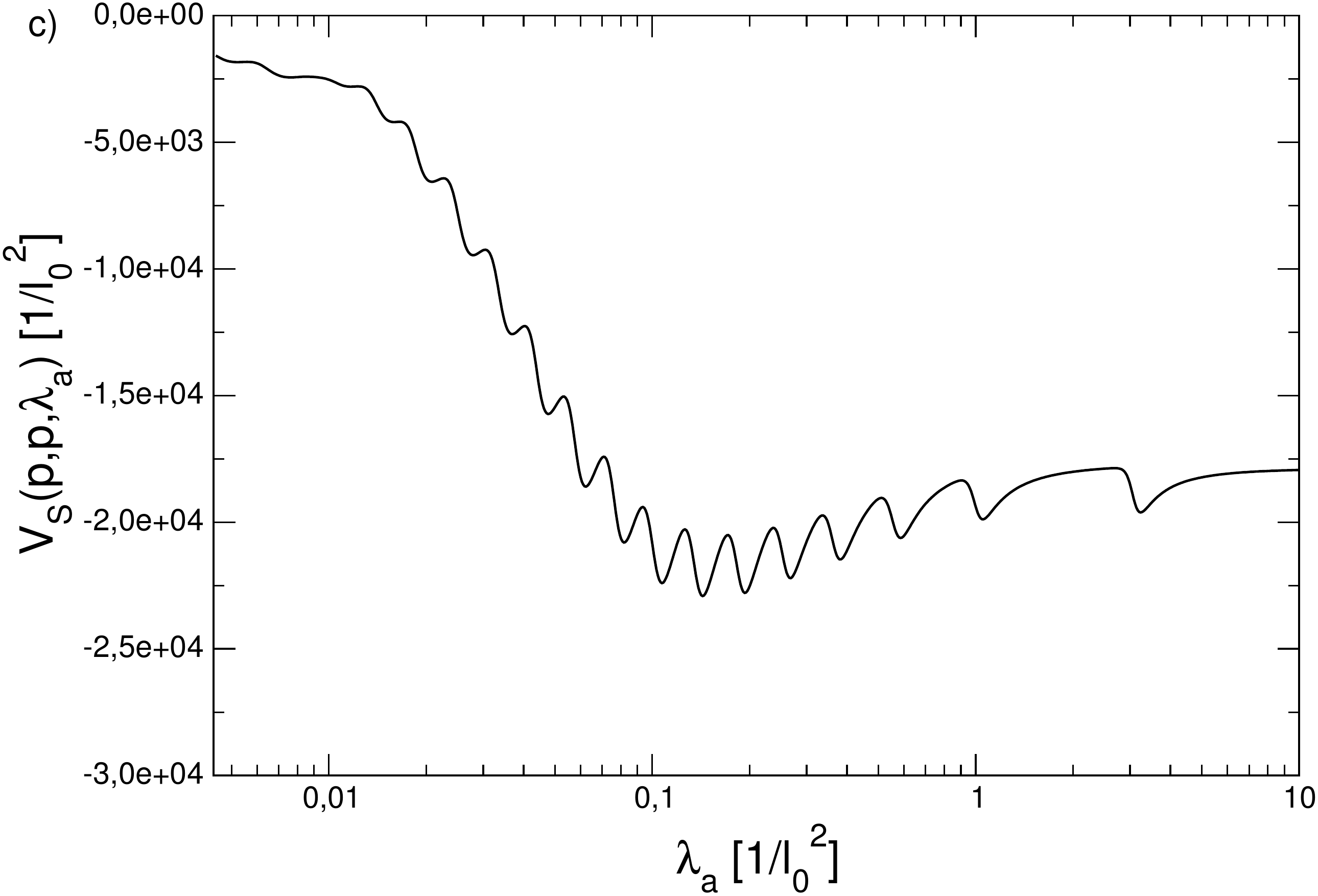}
\end{minipage}
\hfill
\begin{minipage}[hbt]{7.5cm}
	\centering
	\includegraphics[width=7.5cm]{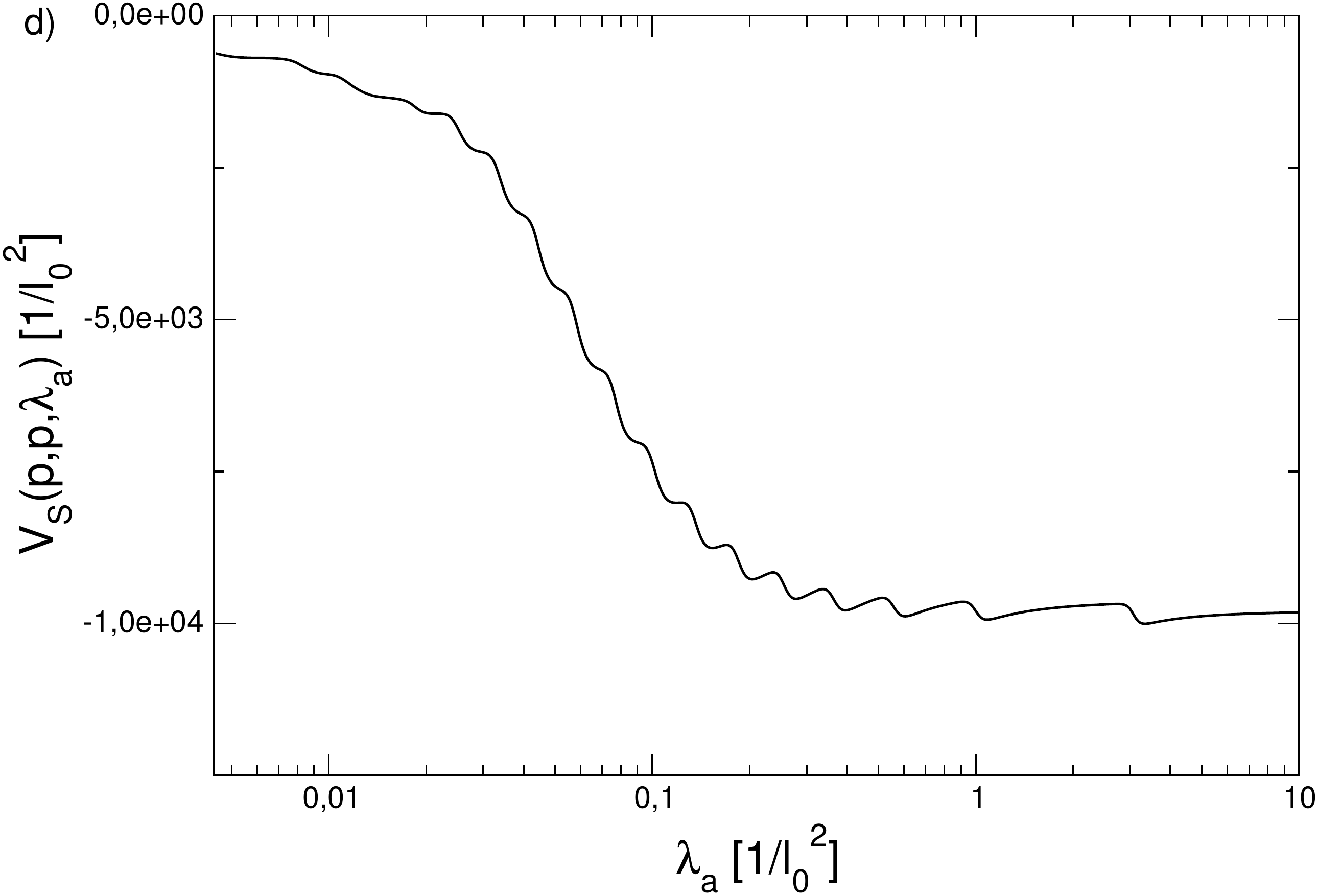}
	\end{minipage} \\
	\caption{Diagonal elements $V(p,p,\lambda_a)$ evolved with the 
inverse generator for four different momenta $p$  with constants $\nu=11$, 
$\Lambda=30 \;l_0^{-1}$ and $\sigma=0.05 \; l_0^{-1} $: 
(a) $p \approx 0.96 \sigma$ , (b)  $p \approx 1.55 \sigma$, 
(c) $p \approx 2.68 \sigma$, (d) $p \approx 4.88 \sigma$.}
	\label{fig:diagonalen}
\end{figure}

We will now elaborate on the appearance of the oscillations on the
diagonal. To this effect, the diagonal elements $V(p,p,\lambda_a)$
are displayed in Fig.~\ref{fig:diagonalen} as a 
function of $\lambda_a$ for four different momenta. The 
clearest oscillations can be extracted in the region $p \approx \sigma$. 
For larger momenta, the amplitudes of the oscillations become smaller, 
cf.~Fig.~\ref{fig:diagonalen}(c) and (d). If $p$ is even further increased, 
the oscillations disappear. Figure~\ref{fig:diagonalen} (a) demonstrates 
that choosing momenta smaller than $\sigma$ leads to
fewer oscillations on the diagonal. 
If $p$ is even further decreased, 
the number of oscillations with large amplitude 
is reduced and the graph resembles the diagonals of potentials 
evolved with the $T$ generator.
This behavior is expected since the inverse and exponential generators
reduce to the standard $T$ generator for small momenta $p$.
We briefly return to Fig.~\ref{fig:diagonalen}(b). For 
smaller $\lambda_a$,  the amplitude of the oscillation 
decreases until no oscillation is visible anymore. The number 
of observable oscillations strongly depends on $\sigma$. 
Typically, they can be seen until $\lambda$ roughly equals
$\sigma$. 

In summary, we find a clear signal of the limit cycle in the $\lambda_a$
dependence of  $V(p,p,\lambda_a)$ for $p\sim \sigma$. In this region,
the evolution of the alternative generators is distinctively different 
from the standard $T$ generator but not heavily suppressed 
as for large $p\gg \sigma$. In practice, many oscillatory features are
compressed in the region $p\sim\sigma$ and thus can be distinguished from 
other effects of the evolution. They can be used
to extract the preferred scaling factor numerically. 

\section{Summary and Outlook}

In this paper, we have investigated the renormalization group
limit cycle of the quantum-mechanical $1/R^2$ potential 
within the similarity renormalization group framework.
We showed that the period of the limit cycle can be extracted 
from the $\lambda_a$ dependence of the diagonal 
elements of the evolved interaction $V(p,p,\lambda_a)$
if the alternative generators in Eq.~(\ref{eq:agen})
with a dimensionful parameter $\sigma$ are used. 
In the region $p\sim \sigma$ sufficiently many oscillatory features
are present, such that a numerical extraction of the preferred
scaling factor is possible.  Here, we have 
determined the period simply by taking averages over the positions of
minima and maxima in $V(p,p,\lambda_a)$. More elaborate 
statistical analysis schemes using Bayesian statistics
can help to improve the extraction 
of the period \cite{Furnstahl:2014xsa}.

In the future, it will be interesting to apply our technique
to the nuclear three-body system in order to derive  the 
limit cycle in the pionless EFT from a more fundamental
chiral interaction with explicit pions. Work in this 
direction is in progress.

{\it Acknowledgements.}\quad 
Useful discussions with Dick Furnstahl 
and Lucas Platter are gratefully acknowledged. This 
work was supported in part by the Helmholtz Association under contract 
HA216/EMMI, by the BMBF (grant 06BN9006), and by the DFG
through SFB 634.

\end{document}